\documentclass[conference,a4paper]{APSIPA2021}
\usepackage{amsmath}
\usepackage{graphicx}
\usepackage{multirow}
\usepackage{threeparttable}
\usepackage[backend=biber,style=ieee,]{biblatex}
\usepackage{tabularray}
\usepackage{array}
\usepackage{multirow}
\usepackage{booktabs}

\usepackage{geometry}
\usepackage{fancyhdr}

\fancypagestyle{firststyle}{
  \fancyhf{}
  \fancyhead[C]{2024 Asia Pacific Signal and Information Processing Association Annual Summit and Conference (APSIPA ASC)}
}

\begin{document}

\title{Detecting Spoof Voices in Asian Non-Native Speech: An Indonesian and Thai Case Study}

\author{
\authorblockN{
Aulia Adila,
Candy Olivia Mawalim, and
Masashi Unoki
}

\authorblockA{
Japan Advanced Institute of Science and Technology, Japan \\
E-mail: \{adila, candylim, unoki\}@jaist.ac.jp}
}

\maketitle
\thispagestyle{firststyle}
\pagestyle{fancy}

\begin{abstract}
This study focuses on building effective spoofing countermeasures (CMs) for non-native speech, specifically targeting Indonesian and Thai speakers. We constructed a dataset comprising both native and non-native speech to facilitate our research. Three key features—MFCC, LFCC, and CQCC—were extracted from the speech data, and three classic machine learning-based classifiers—CatBoost, XGBoost, and GMM—were employed to develop robust spoofing detection systems using the native and combined (native and non-native) speech data. This resulted in two types of CMs: Native and Combined. The performance of these CMs was evaluated on both native and non-native speech datasets. Our findings reveal significant challenges faced by Native CM in handling non-native speech, highlighting the necessity for domain-specific solutions. The proposed method shows improved detection capabilities, demonstrating the importance of incorporating non-native speech data into the training process. This work lays the foundation for more effective spoofing detection systems in diverse linguistic contexts.
\end{abstract}

\section{Introduction}
Over the past few years, voice-based authentication systems have become prominent for verifying identities and recognizing spoken utterances, offering convenience in various scenarios. However, they are vulnerable to spoofing attacks, particularly through logical access (LA) scenarios \cite{ASV2019} involving Text-to-Speech (TTS) and Voice Conversion (VC). While much research has focused on spoofing detection in English and other languages \cite{habla, add}, there is a significant gap in studies addressing non-native accents, particularly within the Asian region. English-speaking accents in these countries present unique challenges, as non-native pronunciations make it difficult for spoofing detection systems to accurately differentiate between bonafide and spoof speech. 
%% can be erased
% TTS generates natural-sounding artificial speech from text, while VC converts one speaker's voice to another's \cite{PresentationAttack}. These techniques can create highly realistic speech, posing significant threats to the security and reliability of these systems. 
%% can be erased
% The latest ASVSpoof 5 challenge \cite{Wang2024_ASVspoof5} has demonstrated new surrogate detection models with adversarial attacks incorporated for the first time, using non-studio quality recordings that introduced a new challenge in the dataset. They built end-to-end systems using RawNet2 \cite{rawnet2} and AASIST \cite{aasist}.

% While much research has focused on spoofing detection in English and other languages \cite{habla, add}, there is a notable gap in studies addressing specific linguistic contexts, such as the diverse languages within the Asian region. English-speaking accents in Asian countries present unique challenges, as non-native accents make it difficult for spoofing detection systems to accurately differentiate between bonafide (genuine) and spoof speech. 

This gap becomes critical in applications such as language proficiency tests, where foreign speakers might attempt to fool automated systems using TTS or VC technologies. In high-stakes scenarios like visa applications or educational admissions, spoofing could manipulate speech to sound more proficient, compromising the test's integrity. Research on non-native speech spoofing is essential to develop robust countermeasures, ensuring the fairness and reliability of systems that evaluate language proficiency and identity verification across diverse linguistic backgrounds.

This study focuses on Indonesia and Thailand due to Indonesia's large multicultural population and Thailand's unique dialect variability and tonal representation. Given the lack of publicly available datasets for these linguistic contexts, we developed an audio dataset with speech from Indonesian and Thai non-native English speakers. Our goal is to establish a spoofing countermeasure (CM) to effectively detect TTS and VC attacks for non-native speech accents and patterns.

The contribution of our work is summarized as follows:
\begin{itemize}

\item to facilitate the development of CMs to handle non-native speech by creating a dataset with non-native speech, addressing the gap in available speech datasets comprising non-native accents and characteristics,

\item to understand the effectiveness on Native CMs in handling Asian non-native speech, and

\item to propose CMs that can distinguish bonafide and spoof speech for both natives and non-natives.

% contribution:
% dataset
% test native CM (current CM) to non-native sets
% exp 1 (build native cm) and 2 (build combined cm)
 
\end{itemize}

In this study, we created a comprehensive dataset consisting of English native and Indonesian-Thai non-native speech to construct the CMs. Furthermore, we investigated how the CMs performed in distinguishing bonafide and spoof speech for both native and non-native speakers. Our findings show that the proposed method significantly improved CM performance in handling non-native speech. Additionally, our CMs utilize common front-end features and back-end classic machine learning (ML)-based classifiers, establishing a foundational baseline for this non-native speech study.

% The remainder of this paper is outlined as follows. Section 2 details the collection and development of our dataset, which includes both bonafide and spoof speech samples. Section 3 provides an overview of our system, covering acoustic feature extraction and the back-end classifier. Section 4 discusses the experimental setup used for system development. Section 5 presents our evaluation results and observations from the dataset. Lastly, Section 6 concludes our findings in this work.

\section{Related Works}
In response to the growing threat of spoofing, the research community launched the ASVspoof challenges\footnote{\url{https://www.asvspoof.org/}}, a biennial initiative aimed at developing CMs to detect fake speech, starting in 2015 \cite{ASV2015} and continuing through 2024 \cite{Wang2024_ASVspoof5}. As these types of attacks have increased through the challenge editions, TTS and VC attacks have been included since the first edition in 2015.

Several CM systems have been developed to detect audio spoofing, primarily focusing on hand-crafted features that can effectively capture discriminative patterns of artifacts \cite{AudioAntiSpoofingSurvey}. Early research mainly used conventional classifiers, such as ML-based classifiers. 

A study on synthetic speech detection in 2015 found that dynamic Linear Frequency Cepstral Coefficients (LFCC) with a GMM classifier performed best on ASVspoof evaluation sets \cite{FeatureComparison2015}. In ASVspoof 2017 \cite{ASV2017}, the best system in the previous challenge was used as the baseline system. It was built using a common GMM back-end classifier with constant Q cepstral coefficients (CQCC). In ASVspoof 2019 \cite{ASV2019}, the top-performing systems employed LFCC features with a light convolutional neural network (LCNN). However, other systems based on standard cepstral features and GMM-based classifiers were only slightly behind in performance. In ASVspoof 2021 \cite{ASV2021}, most systems operated using short-term spectral features or raw waveforms, utilizing ensemble systems and popular convolutional networks. These efforts primarily addressed spoofing in native language domains, leaving the non-native speech domain largely unexplored.
Speech corpora are essential for CM systems, and publicly available datasets with a large number of spoofing attacks have emerged to overcome data bottlenecks. These datasets, which include both bonafide and spoof speech generated by LA algorithms (TTS and/or VC), facilitate the evaluation and benchmarking of different systems. Notable datasets include ASVspoof2015 \cite{ASV2015}, ASVspoof2019-LA \cite{ASV2019_dataset}, ASVspoof2021-LA \cite{ASV2021}, and the latest ASVspoof2024 \cite{Wang2024_ASVspoof5}. Other available datasets include ADD2022-LF \cite{add} and Latin-American Voice Anti-spoofing \cite{habla}. However, datasets representing non-native English speech are scarce, highlighting the need for research focused on spoofing detection tailored to non-native accents.

%% added
While some non-native speech datasets exist \cite{speechocean, ISLE_nonnative}, such as those designed for automatic speech recognition (ASR) and language learning, they are not specifically tailored for spoofing detection tasks. These datasets typically focus on improving speech recognition for non-native speakers or assessing language proficiency rather than detecting synthetic or manipulated speech. The absence of non-native speech datasets for spoofing detection further underscores the importance of developing and expanding resources to support more effective countermeasures in diverse linguistic environments.

\section{Dataset Collection}
The dataset used in this study consists of English native and non-native speech, spoken by Indonesian and Thai speakers. The dataset is split into training, validation, and testing sets with the ratio of 70-10-20 to ensure robust evaluation, with no overlap of speakers between the splits to maintain the integrity of the results. Table \ref{tab:datacollection} summarizes our ENIT Dataset\footnote{ENIT Dataset: English Native and Indonesian-Thai Non-Native Speech}.

\subsection{Bonafide Data}
Our native English dataset consists of 7,990 utterances sourced from the training sets of ASVspoof 5 \cite{Wang2024_ASVspoof5}, which is derived from the English-language subset of the Multilingual Librispeech (MLS) dataset. The MLS dataset is a large multilingual corpus based on LibriVox audiobooks \cite{MLSData}, featuring non-studio-quality recordings. This dataset includes recordings from over 4,000 speakers using various devices. As this dataset has a balanced number of speakers, we randomly selected 4,000 utterances from the total of 80 males and 80 females speakers and eliminated 10 utterances randomly to ensure the balance with our non-native English dataset. 

\begin{table}[]
\caption{ENIT Dataset: English Native and Indonesian-Thai \newline Non-Native Speech}
\centering
\begin{tabular}{@{}ccccccc@{}}
\toprule
\multirow{2}{*}{Type} &
\multirow{2}{*}{Sets} &
\multicolumn{2}{c}{\# unique spk.} &
\multicolumn{2}{c}{\# utterances} &
\# spoof \\ \cmidrule(lr){3-6}
& & Male & Female & Bonafide & Spoof & attack \\ \midrule
\multirow{3}{*}{Native} & Train. & 56 & 56 & 5,590 & 33,540 & 8 \\
& Dev. & 8 & 8 & 800 & 4,800 & 8 \\
& Eval. & 16 & 16 & 1,600 & 9,600 & 8 \\ \midrule
\multirow{3}{*}{\begin{tabular}[c]{@{}c@{}}Non-\\ native\end{tabular}} &
Train. &
10 / 12 &
4 / 5 &
5,696 &
33,164 &
3 \\
& Dev. & 1 / 1 & 1 / 2 & 542 & 4,275 & 3 \\
& Eval. & 3 / 3 & 2 / 3 & 1,752 & 10,501 & 3 \\ \bottomrule
\end{tabular}
\label{tab:datacollection}
\end{table}

The English non-native speech dataset consisted of 7,990 utterances recorded from 21 speakers, including 10 Indonesian speakers (7 males and 3 females) and 11 Thai speakers (7 males and 4 females). These speakers read articles sourced from online English newsletters covering unbiased topics such as health, astronomy, engineering, and technology. We selected similar news topics in Indonesian, English, and Thai, ensuring they did not express hate speech, violence, or harassment to maintain impartiality. Each news article was then segmented into sentences or sub-sentences with 5 to 20 words per utterance to ensure readability.

The recordings were conducted in a soundproof room using four different recording devices: a condenser microphone, a dynamic microphone, a mobile phone, and a laptop computer, without significant channel or background noise effects. All speech was recorded in a read style, with the speaking pace adjusted to each speaker, while still maintaining the naturalness of spontaneous speech. Audio post-processing included practical trim voice editing using Audacity\footnote{Audacity (ver. 3.4.2) is an open source software for recording and editing sounds.} to separate the recordings by utterances.

\subsection{Spoof Data}
We constructed a corpus of 47,940 utterances by selecting a subset from the ASVspoof 5 training data \cite{Wang2024_ASVspoof5}. The selected utterances are from the same 80 males and 80 females speakers used in the bonafide data. The number of spoof utterances is six times larger than the number of bonafide utterances. Table \ref{tab:datacollection} specifies the number of unique speakers of bonafide and spoof speech on the left and right side of the separator `/', respectively. If the number of speakers is the same for both bonafide and spoof speech, we do not use the separator.

To generate spoof data for non-native English speech, we employed TTS, VC, and synthesis techniques. These methods can be broadly categorized into three distinct approaches.

\begin{figure*}[t]
\begin{center}
\includegraphics[width=.65\linewidth]{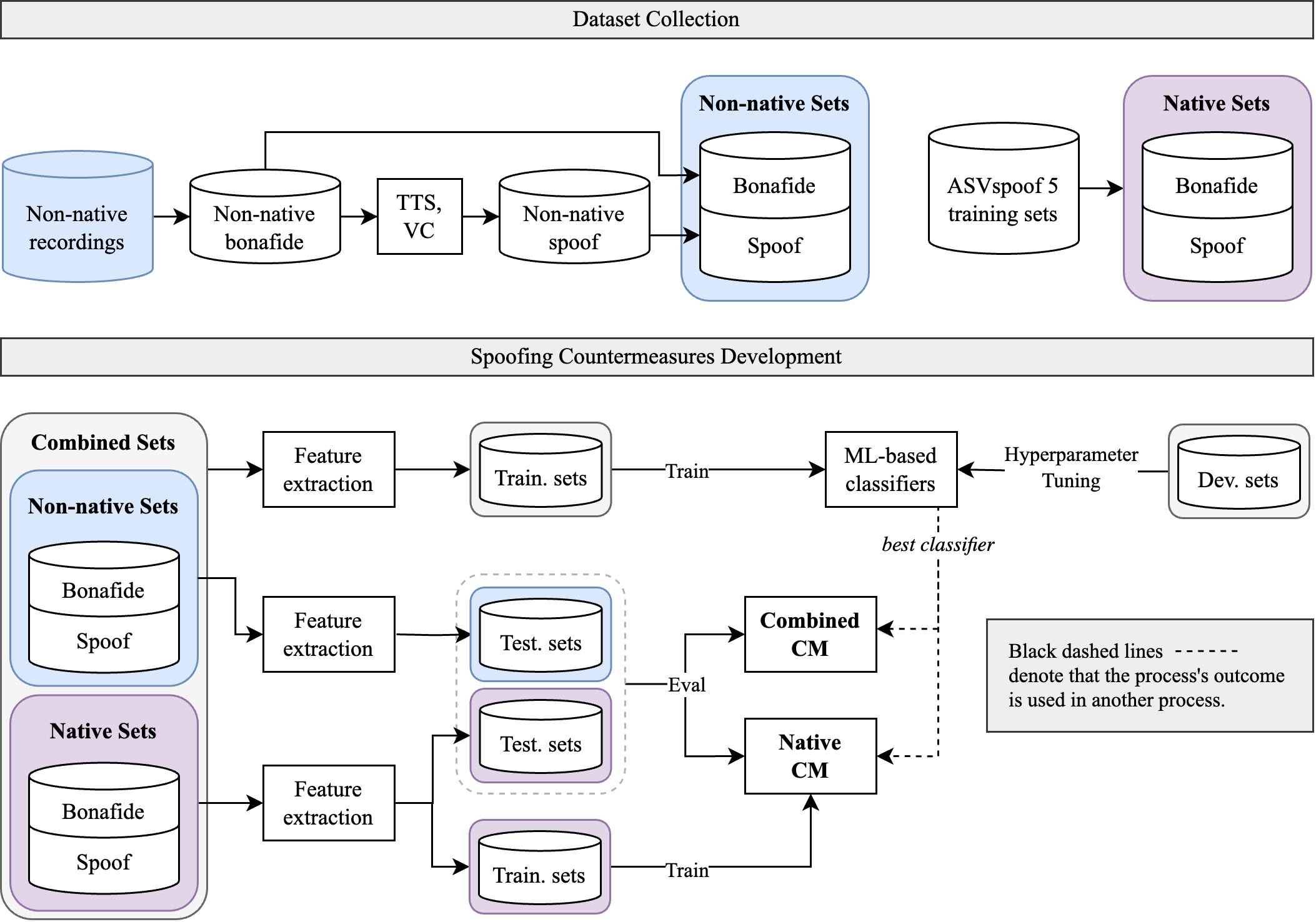}
% \vspace{-5mm}
\caption{Block diagrams of our proposed method for building spoofing countermeasures}
\label{fig:pipeline}
\vspace{-3mm}
\end{center}
\end{figure*} 

\subsubsection{SpeechT5 \cite{Ao2022_SpeechT5}}
SpeechT5 is a model that unifies speech and text processing under a single framework. Inspired by the success of T5 in the text domain \cite{RaffelSRLNMZLL20_T5}, SpeechT5 employs an encoder-decoder architecture to learn shared representations for both modalities. It incorporates specialized pre- and post-processing networks to handle the unique characteristics of speech and text, respectively.

Through extensive self-supervised pre-training on large-scale unlabeled speech and text data, SpeechT5 acquires a deep understanding of both modalities. This versatility empowers it to excel in a wide range of tasks, including automatic speech recognition, speech synthesis, speech translation, VC, speech enhancement, and speaker identification.

Approximately 5,200 synthetic utterances were generated using the SpeechT5 model and x-vector speaker embeddings \cite{SnyderGSPK18_xvector}. To manipulate acoustic and linguistic characteristics, we varied both x-vector and bottleneck features (representing linguistic information). Roughly 2,000 utterances were created using x-vector and bottleneck features extracted from our collected bonafide non-native speaker data. In another set of 1,000 utterances, x-vectors from native speakers in the CMU ARCTIC dataset \cite{KominekB04a_CMU_Arctic} were combined with bottleneck features from our non-native speakers. The remaining utterances were synthesized by pairing x-vectors from our non-native speakers with bottleneck features from the VCTK dataset \cite{yamagishi2019vctk}.

\subsubsection{FreeVC \cite{LiTX23a_FreeVC}}
We utilized FreeVC, a text-free VC system capable of transforming a speaker's voice with only a single reference audio sample. The model builds upon the ViTS architecture \cite{KimKS21_ViTS} but departs from traditional text-based approaches by learning to disentangle content information directly from raw audio waveforms. FreeVC extracts speaker-independent features that are subsequently processed through a bottleneck layer to isolate content representations using WavLM \cite{ChenWCWLCLKYXWZ22_WavLM}. To further enhance the model's ability to separate content from speaker identity, spectrogram resizing is employed as a data augmentation technique.

Central to the one-shot conversion capability is a speaker encoder that extracts speaker-specific characteristics. We experiment with two configurations of the speaker encoder: FreeVC-s, which utilizes a non-pretrained encoder, and FreeVC, which benefits from a pre-trained encoder. This comparative analysis allows us to assess the impact of pre-training on the overall performance of the VC system.

\subsubsection{WORLD \cite{MoriseYO16_WORLD}}
We also employed the WORLD, a vocoder-based speech synthesis system, to generate synthetic spoof data. This system is widely recognized for its exceptional performance in speech analysis, manipulation, and synthesis. It analyzes speech into three parts: fundamental frequency (F0), aperiodicity, and spectral envelope. 

To further refine the spectral envelope estimation process, we integrated the CheapTrick algorithm \cite{Morise15_CheapTrick} into our pipeline. This method leverages F0-adaptive windowing to enhance spectral resolution. By smoothing the power spectrum and applying spectral recovery techniques in the quefrency domain, CheapTrick better estimates the spectral envelope. This improved accuracy contributes significantly to the overall quality of the synthesized speech. We manipulated speed and F0 with random parameters during the speech synthesis process. This randomization helps to create a more diverse and challenging dataset for subsequent anti-spoofing models.

\section{Proposed Method}
A block diagram of our proposed method in building the spoofing CMs using our datasets is shown in Figure \ref{fig:pipeline}. To establish a foundation for our newly developed dataset, which represents the characteristics of non-native speech, we employed hand-crafted features and classic ML-based classifiers. 
% This approach allows us to develop a baseline for our dataset. 

\subsection{Feature extraction}
The front-end features are mostly derived from the magnitude or power spectrum, encompassing both short-term and long-term magnitude spectral features \cite{AudioAntiSpoofingSurvey}, to effectively capture the essential characteristics of speech signals, including formant structures and energy distribution across different frequency bands. These features are typically obtained from MFCC \cite{1163420}, LFCC \cite{6712706}, and CQCC \cite{tak2020explainabilitystudyconstantq}. 

LFCC is widely used in speaker recognition and has demonstrated strong performance in spoofing detection. MFCC has been extensively explored for accent classification tasks, which closely relate to this study on non-native speech. CQCC has also proven effective in detecting audio spoofing by providing higher frequency resolution at lower frequencies and higher temporal resolution at higher frequencies. This study aims to investigate the effectiveness of LFCC, MFCC, and CQCC in detecting spoofing attacks in non-native English speech, focusing on Indonesian and Thai accented speakers.

\subsection{Classifiers}
We use classic ML-based classifiers to build the baseline spoofing CMs. To determine the most effective classifier, we conducted model selection on several widely used ML-based classifiers and evaluated their performance on our development sets. On the basis of our experiments, we identified the two best classifiers for the features used (MFCC, LFCC, and CQCC): CatBoost \cite{prokhorenkova2019catboostunbiasedboostingcategorical} and XGBoost \cite{Chen_2016}.

CatBoost is a type of binary decision trees-based gradient-boosting predictor that has proven effective in dealing with categorical data \cite{dorogush2018catboostgradientboostingcategorical}. Categorical data consists of distinct values that cannot be compared directly. In this study, we deal with binary categorical data.

Extreme Gradient Boosting, or XGBoost, is a gradient-boosted tree algorithm for supervised learning. Gradient-boosted decision trees excel in learning from noisy data and have achieved cutting-edge results. XGBoost is a gradient-boosting implementation that optimizes computing performance to provide a scalable solution.

Additionally, we included the GMM-based classifier as our baseline, which has been a fundamental method in the ASVspoof challenge series. GMM assumes that data points belong to a mixture of a finite Gaussian distribution \cite{scikit-learn}. It is commonly used in anti-spoofing systems. There, separate GMMs are learned for each bonafide and spoof dataset. The classification of new input is predicted by calculating the ratio of its log-likelihood of belonging to bonafide and spoof GMMs.

\section{Experimental Settings}
As the currently available speech corpora primarily consist of native speech, we aim to assess the performance of CMs built upon these datasets in handling non-native speech. We refer to this CM as Native CM. Subsequently, we developed a spoofing CM specifically for non-native speech by combining the current publicly available datasets, which are predominantly native speech, with our non-native datasets. We refer to this CM as Combined CM. Our experiment pipeline is illustrated in Fig. \ref{fig:pipeline}.

\subsection{Experiment 1: Assessing the Native CM on Non-Native Speech}
First, we built the Native CM by training our selected models (GMM, CatBoost, and XGBoost) using the extracted features (MFCC, LFCC, and CQCC) on the native speech training sets. The features were extracted using the Matlab toolbox and Smileslab implementations \cite{Smileslab}. We set the random seed to 42 for all classifiers and adjusted the hyperparameters as detailed in Subsection \ref{subsec:hyperparameter}. Each model was then evaluated on both native and non-native speech evaluation sets, using the specified evaluation metrics. This initial assessment on native speech helps us understand the model's performance in a typical native speech context.

\subsection{Experiment 2: Assessing the Combined CM on Non-Native Speech}
With regard to the Combined CM, we trained our selected models (GMM, CatBoost, and XGBoost) using the extracted features (MFCC, LFCC, and CQCC) on the combined native and non-native speech training sets. We set the same hyperparameter tuning as experiment 1 to ensure fair comparison. We also used the same evaluation protocol to facilitate the common assessment and benchmarking of both CMs.

\subsection{Hyperparameter settings} \label{subsec:hyperparameter}
We set different hyperparameters for each classifier on the basis of its architecture as implemented in the scikit-learn library \cite{scikit-learn}. For the GMM classifier, we used two mixture components, a full covariance type, the k-means method for initializing the weights, and a maximum of 100 iterations. For CatBoost, we set the verbose parameter to 0. For XGBoost, we used logarithmic loss as the evaluation metric. All other hyperparameters were kept at their default configurations.

\begin{table}[]
\caption{CMs performance evaluation result in Experiments 1 and 2}
\resizebox{\columnwidth}{!}{
\begin{tabular}{|c|c|c|cc|cc|}
\hline
\multirow{2}{*}{Exp.} & \multirow{2}{*}{Feat.} & \multirow{2}{*}{Classifier} & \multicolumn{2}{c|}{Non-native} & \multicolumn{2}{c|}{Native} \\ \cline{4-7} 
& & & \multicolumn{1}{l|}{minDCF} & \multicolumn{1}{l|}{EER (\%)} & \multicolumn{1}{l|}{minDCF} & \multicolumn{1}{l|}{EER (\%)} \\ \hline \hline
\multirow{9}{*}{1} & \multirow{3}{*}{MFCC} & CatBoost & \multicolumn{1}{c|}{0.81} & 41.54 & \multicolumn{1}{c|}{\textbf{0.17}} & \textbf{6.88} \\ \cline{3-7} 
& & XGBoost & \multicolumn{1}{c|}{0.83} & 42.58 & \multicolumn{1}{c|}{0.18} & 7.48 \\ \cline{3-7} 
& & GMM & \multicolumn{1}{c|}{1.00} & 100 & \multicolumn{1}{c|}{1.00} & 100 \\ \cline{2-7} 
& \multirow{3}{*}{LFCC} & CatBoost & \multicolumn{1}{c|}{0.95} & 40.07 & \multicolumn{1}{c|}{0.24} & 10.19 \\ \cline{3-7} 
& & XGBoost & \multicolumn{1}{c|}{0.98} & 40.36 & \multicolumn{1}{c|}{0.26} & 10.56 \\ \cline{3-7} 
& & GMM & \multicolumn{1}{c|}{1.00} & 62.84 & \multicolumn{1}{c|}{0.83} & 33.89 \\ \cline{2-7} 
& \multirow{3}{*}{CQCC} & CatBoost & \multicolumn{1}{c|}{0.84} & 38.11 & \multicolumn{1}{c|}{0.29} & 12.43 \\ \cline{3-7} 
& & XGBoost & \multicolumn{1}{c|}{0.79} & 35.57 & \multicolumn{1}{c|}{0.28} & 12.56 \\ \cline{3-7} 
& & GMM & \multicolumn{1}{c|}{1.00} & 76.02 & \multicolumn{1}{c|}{0.97} & 47.69 \\ \hline \hline
\multirow{9}{*}{2} & \multirow{3}{*}{MFCC} & CatBoost & \multicolumn{1}{c|}{0.33} & 12.66 & \multicolumn{1}{c|}{0.19} & 7.81 \\ \cline{3-7} 
& & XGBoost & \multicolumn{1}{c|}{0.38} & 13.98 & \multicolumn{1}{c|}{0.21} & 8.08 \\ \cline{3-7} 
& & GMM & \multicolumn{1}{c|}{0.96} & 46.58 & \multicolumn{1}{c|}{1.00} & 46.18 \\ \cline{2-7} 
& \multirow{3}{*}{LFCC} & CatBoost & \multicolumn{1}{c|}{0.27} & 10.90 & \multicolumn{1}{c|}{0.31} & 13.32 \\ \cline{3-7} 
& & XGBoost & \multicolumn{1}{c|}{0.25} & 10.34 & \multicolumn{1}{c|}{0.33} & 13.76 \\ \cline{3-7} 
& & GMM & \multicolumn{1}{c|}{0.98} & 53.31 & \multicolumn{1}{c|}{1.00} & 46.13 \\ \cline{2-7} 
& \multirow{3}{*}{CQCC} & CatBoost & \multicolumn{1}{c|}{\textbf{0.19}} & \textbf{8.56} & \multicolumn{1}{c|}{0.37} & 15.06 \\ \cline{3-7} 
& & XGBoost & \multicolumn{1}{c|}{0.21} & 9.63 & \multicolumn{1}{c|}{0.36} & 14.75 \\ \cline{3-7} 
& & GMM & \multicolumn{1}{c|}{0.97} & 41.55 & \multicolumn{1}{c|}{0.99} & 91.89 \\ \hline
\end{tabular}}
\label{tab:result}
\vspace{-4mm}
\end{table}

\subsection{Evaluation Metrics}
In this study, we used minimum detection cost function (minDCF) and equal error rate (EER) adopted from the latest ASVspoof 5 challenge, specifically on track 1: stand-alone spoofing and speech deepfake detection \cite{Wang2024_ASVspoof5}. The spoofing detection employed in the challenge is built upon the normalized detection cost function (DCF), defined as follows:

\begin{equation}\label{eq:normDCF}
\text{DCF}'(\tau_{\text{cm}}) = \beta \cdot P_{\text{cm, miss}}(\tau_{\text{cm}}) + P_{\text{cm, fa}}(\tau_{\text{cm}})
\end{equation}
where
\begin{equation}\label{eq:b}
\beta = \frac{C_{\text{miss}} \cdot (1 - \pi_{\text{spf}})}{C_{\text{fa}} \cdot \pi_{\text{spf}}}
\end{equation}

In Eq. (\ref{eq:normDCF}), $P_{\text{cm, miss}}(\tau_{\text{cm}})$ and $P_{\text{cm, fa}}(\tau_{\text{cm}})$ are the empirical miss rate for bonafide utterances and false alarms for spoof utterances, respectively. Both are regarded as a function of the detection threshold $\tau_{\text{cm}}$. The constant $\beta$ In Eq. (\ref{eq:b}) is calculated using miss rate cost $C_{\text{miss}}$, false alarm cost $C_{\text{fa}}$, and detection threshold $\tau_{\text{cm}}$ with the values set as follows: $C_{\text{miss}} = 1$, $C_{\text{fa}} = 10$, and $\pi_{\text{spf}} = 0.05$. Therefore, the value of $\beta$ is approximately $1.90$.

Furthermore, Eqs. (\ref{eq:normDCF}) and (\ref{eq:b}) are used to compute the minimum DCF (minDCF), which measures the CMs performance by using the threshold set on the basis of ground-truth where the lower value indicates a better performance. This metric is calculated as follows:

\begin{equation}
\text{minDCF} = \min_{\tau_{\text{cm}}} \text{DCF}'(\tau_{\text{cm}}) \quad 
\end{equation}

The final evaluation metric used is the EER, one of most widely used metrics for audio spoofing CMs. It represents the CM threshold where the false acceptance rate equals the false rejection rate. A lower EER value signifies better performance. 

\section{Results}
As seen in Table \ref{tab:result}, the GMM classifier has comparably high minDCF and EER values on the native speech evaluation sets. It also reached the maximum possible DCF and EER values of $100$\% on MFCC features, indicating extremely poor performance as it is entirely ineffective at distinguishing between bonafide and spoof speech in both non-native and native contexts. The lowest EER for the GMM classifier was achieved with LFCC features, which is expected as this system is considered the baseline in the ASVspoof challenge.

While the CatBoost and XGBoost classifiers achieved competitive results on native datasets, they still obtained minDCF values greater than $0.79$ and EER values higher than $38$\% on non-native datasets. All the CMs in experiment 1 on average declined around $0.44$ in minDCF and $19\%$ in EER relative to their performance on native datasets. These results indicate a significant lack of non-native speech handling capability in Native CMs, demonstrating the inefficacy of the pre-existing speech spoofing CMs in distinguishing between bonafide and spoof speech in non-native contexts.

In experiment 2, all Combined CMs significantly improved when predicting the non-native datasets, with an average relative improvement of around $0.41$ in minDCF and around $30$\% in EER. However, this experimental setting slightly worsened the CMs' ability to detect native speech, as indicated by an average of total increase on minDCF and EER  around $0.06$ and $1.7$\%, respectively. This behavior may be due to the increased variability introduced by the non-native speech data within the training sets, making it more challenging for the model to generalize well to the specific characteristics of native speech alone, as it now has to account for a broader range of speech patterns.

Among all the CMs we built, CQCC feature extraction with CatBoost classifier performed the best when evaluated on non-native datasets. These ensemble models tended to perform better due to their boosting algorithm, which iteratively builds an ensemble by training each new model to correct the errors of the previous ones.  

Meanwhile, the GMM classifier, which is often used as a baseline model, performed worse than other classifiers. In most cases, the GMM classifier predicted the input speech as a spoof. Despite reducing EER by up to $50$\% after being trained using the combined native and non-native datasets, its performance as a CM is not satisfactory. 

\section{Conclusions}
In this study, we constructed spoofing countermeasures (CMs) to handle non-native speech using different acoustic features (MFCC, LFCC, and CQCC) and classic machine learning-based classifiers (CatBoost, XGBoost, and GMM). We also developed the ENIT dataset, which includes speech from Indonesian and Thai non-native English speakers. We conducted two experiments to assess the current CMs (Native CM) in handling non-native speech, as well as the baseline we constructed (Combined CM). 

Our findings show that the Combined CMs is better at detecting bonafide and spoof non-native speech than the Native CMs, improving minDCF and EER scores around $0.41$ and $30$\% on average, respectively. Moreover, we also found that CatBoost and XGBoost obtained competitive results using all three features, while GMM performed the worst in all experiments. The future work aims to expand datasets for Asian non-native speech and develop robust CMs using deep learning, focusing on accent and speech characteristics.

% In our future work, we will expand the datasets to be able to capture more non-native speech from the Asian region, as well as building more accurate and robust CMs using deep learning, with a focus on accent and other speech characteristics. 

\section*{Acknowledgments}
This work was supported by a Grant-in-Aid for the Promotion of Joint International Research (Fostering Joint International Research (B)) (20KK0233), Grant-in-Aid for Challenging Research (Exploratory) (23K18491), and the SCAT Foundation. This work was a part of the ASEAN IVO project titled ‘Spoof Detection for Automatic Speaker Verification’(www.nict.go.jp/en/asean\_ivo).

% \bibliographystyle{IEEEtran}
% \bibliography{mybib}

\printbibliography

\end{document}